\documentclass[prd,twocolumn,floatfix,amsmath,nofootinbib,amssymb,floatfix]{revtex4}
\usepackage{graphicx,color,dcolumn,booktabs,bm,multirow}
\usepackage{longtable,lscape}
\usepackage{txfonts}
\usepackage{overpic}
\usepackage{amssymb}
\usepackage{array}
\usepackage{indentfirst}
\usepackage{feynmf}   
\usepackage{slashed}  
\usepackage{cases}
\usepackage{color}
\usepackage{multirow}
\usepackage{epstopdf}
\usepackage[utf8]{inputenc}
\usepackage{graphicx,color,dcolumn,booktabs,bm}
\usepackage[colorlinks,
citecolor=blue,
anchorcolor=red,
menucolor=red,
linkcolor=red,
filecolor=red,
runcolor=red,
urlcolor=blue,
frenchlinks=red]{hyperref}
\usepackage{float}

\begin{document}
	
	\title{$d_{N \Omega}$ production in $\Omega d$ scattering process}
\author{Quan-Yun Guo$^{1}$}\email{guoquanyun@seu.edu.cn}
\author{Jing Liu$^{2}$}\email{liujing@hue.edu.cn}
\author{Dian-Yong Chen$^{1,3}$\footnote{Corresponding author}}\email{chendy@seu.edu.cn}
\affiliation{$^1$ School of Physics, Southeast University, Nanjing 210094, People's Republic of China}
\affiliation{$^2$ School of Physics and Mechanical Electrical and Engineering, Hubei University of Education, Wuhan 430205, China}
\affiliation{$^3$ Lanzhou Center for Theoretical Physics, Lanzhou University, Lanzhou 730000, China}
	\date{\today}

\begin{abstract}
In the present work, we propose to investigate the production of $d_{N \Omega}$ in the $\Omega^{-} d \rightarrow p d_{N \Omega}^-$ process by utilizing an effective Lagrangian approach, where $d_{N \Omega}$ is identified as $N\Omega$ bound state with the binding energy $E_{b}=2.46$ MeV. Experimentally, the J-PARC hadron facility proposed to investigate the $K^{-}p \rightarrow \Omega^{-} \bar{K}^{(*)0} K^{+}$ process, which is expected to yield an $\Omega$ beam with the momentum of approximately 3 GeV. Additionally, theoretical studies of the $\psi(2S) \rightarrow \Omega^{-} \bar{\Omega}^{+}$ process at BESIII provided an $\Omega$ beam with the momentum of 774 MeV. Considering these two potential $\Omega$ beam sources, our estimations show that for the $\Omega^{-} d \rightarrow p d_{N \Omega}^-$ process, the cross sections are $\Big(329.7^{+26.9}_{-49.6}\Big)$ $\mu$b, $\Big(174.0^{+26.5}_{-38.2}\Big)$ $\mu$b, $\Big(16.9^{+7.4}_{-7.7}\Big)$ $\mu$b, and $\Big(2.0^{+1.8}_{-1.4}\Big)$ $\mu$b at $P_{\Omega} =$ 0.7, 0.9, 2.0, and 4.0 GeV, respectively, where the central values are estimated with $\Lambda_{r}=1.0$ GeV, and the errors come from the variation of $\Lambda_{r}$ from 0.8 to 1.2 GeV. We also estimate the differential cross sections, which reach the maximum at the forward angle limit. In addition, since the $d_{N \Omega}$ dibaryon predominantly decays into $\Xi \Lambda$. Therefore, we further investigate the $\Omega^{-} d \rightarrow p \Xi^- \Lambda$ process and estimate the relevant cross sections. It is expected that the present estimations can be tested by further experimental measurements at J-PARC and STCF in the future.
	\end{abstract}

\maketitle

\section{Introduction}
\label{sec:Introduction}
The exploration of multiquark candidates has become a significant area of hadron physics over the past two decades. Numerous potential exotic candidates have been observed by the experimental collaborations, such as  Belle/Belle\uppercase\expandafter{\romannumeral2}, BES\uppercase\expandafter{\romannumeral3}, LHCb. (see Ref.~\cite{Belle:2003nnu, Belle:2009and, BESIII:2013ris, Belle:2013yex, BESIII:2013qmu, Belle:2014nuw, LHCb:2019kea, LHCb:2020tqd, LHCb:2021vvq, Chen:2016qju, Guo:2017jvc, Liu:2019zoy} for representative examples). Different theoretical interpretations for these exotic candidates include hadronic molecular states, compact multiquark states, hybrids and dynamics effect, have been proposed.

In the dibaryon family, the typical example is deuteron (denoted as $d$) composed of one proton and one neutron with the weak binding energy $E_{b} =2.22$ MeV, which was discovered in 1931. In addition, the possible dibaryon, $d^{\ast}(2380)$, was studied in the $pn \rightarrow d \pi^{0} \pi^{0}$ process within the energy region of the ABC effect by the CELSIUS/WASA-at-COSY Collaboration in 2009~\cite{Bashkanov:2008ih}, while the mass and width of $d^{\ast}(2380)$ were measured to be approximately 2.36 GeV and 80 MeV, respectively. The subsequent analysis confirmed the existence of $d^{\ast}(2380)$ resonance in the $pn \rightarrow d \pi^{0} \pi^{0}$, $pn \rightarrow d \pi^{+} \pi^{-}$, and $pp \rightarrow d \pi^{+} \pi^{0}$ processes~\cite{WASA-at-COSY:2011bjg, WASA-at-COSY:2012seb}. Later, using a polarized deuteron beam impinged on the hydrogen pellet target and considering the quasifree process, $d^{\ast}(2380)$ was observed in the double-pionic fusion channels $d \pi^{0} \pi^{0}$ and $d \pi^{+} \pi^{-}$ of the $np$ scattering process by the WASA-at-COSY Collaboration~\cite{WASA-at-COSY:2014dmv}, while the basic parameters of $d^{\ast}(2380)$ were determined to be $I(J^{P})=0(3^{+})$, $M \approx 2380$ MeV, and $\Gamma \approx 70$ MeV. Moreover, using the WASA detector setup at COSY, $d^{\ast}(2380)$ was also studied in the $pn \rightarrow pp \pi^{0} \pi^{-}$ and $pn \rightarrow pp \pi^{0} \pi^{0}$ reactions~\cite{WASA-at-COSY:2013fzt, WASA-at-COSY:2014qkg}. In 2017, using the FOREST detector at ELPH~\cite{Ishikawa:2016yiq}, the upper limit of the cross section for the $\gamma d \rightarrow d \pi^{0} \pi^{0}$ process was measured to be  0.034 $\mu$b at $W_{\gamma d}=2.37$ GeV. Theoretically, one can find the prosperity of the dibaryon interpretations of $d^{\ast}(2380)$, including $N \Delta$ and $\Delta \Delta$~\cite{Bashkanov:2013cla, Gal:2013dca, Huang:2013nba, Huang:2014kja, Gal:2016bhp, Dong:2017geu, Dong:2017mio, Dong:2018emq, Lu:2017uey, Vidana:2017qey, Lu:2018gtk, Shi:2019dpo, Kim:2020rwn, Lu:2020qme, Beiming:2021bkj, Brambilla:2022ura}.

In addition to $d^{\ast}(2380)$ resonance, the potential nucleon-hyperon dibaryon, $d_{N \Omega}$, was also studied by various models. Using the chiral SU(3) quark model, the authors in Ref.~\cite{Li:1999bc} investigated the $S-$wave bound state $N\Omega$, and the results showed that the binding energy of $N\Omega$ state ranged from 3.5 to 12.7 MeV. Later, the authors in Ref.~\cite{Dai:2006dgq} studied the structures of $N\Omega$ and $\Delta \Omega$ states with strangeness $S=-3$ in the chiral SU(3) quark model, their results indicated that the $N\Omega$ and $\Delta \Omega$ dibaryons were weakly bound states when the short range part of the quark-quark interaction was controlled by the vector meson exchanges. Using the quark delocalization color screening model~\cite{Pang:2003ty}, the binding energy of $N\Omega$ state was estimated to be 62 MeV. In Ref.~\cite{Chen:2021hxs}, the authors investigated the mass spectra of the $N\Omega$ dibaryon in the $^{5}S_{2}$ channel with $J^{P}=2^{+}$ by using the QCD sum rules, and the results indicated that the binding energy of $N\Omega$ state was about 21 MeV. Using the baryon–baryon interaction model, the authors in Ref.~\cite{Sekihara:2018tsb} investigated the origin of the strong attraction in the $N\Omega (^{5}S_{2})$ interaction, their results showed that the binding energy and width of $N\Omega$ state were 0.1 MeV and 1.5 MeV, respectively. In addition, using the effective Lagrangian approach, the authors in Ref.~\cite{Xiao:2020alj} investigated the possible strong two-body decay modes $\Xi \Lambda$ and $\Xi \Sigma$ of $d_{N\Omega}$ in the hadronic molecular frame, and the estimations indicated that the $d_{N\Omega} \rightarrow \Xi \Lambda$ mode is dominant with a branching fraction around $90\%$. 

Recently, Lattice QCD estimations have also provided important constraints on the binding energy of the $N\Omega$ state. The HAL QCD Collaboration studied the properties of $N\Omega$ state in (2+1)-flavor lattice QCD, where the masses of the $ud$ and $s$ quarks corresponded to $m_{\pi}=875$ MeV and $m_{K}=916$ MeV, respectively. The binding energy of $N\Omega$ state was estimated to be $18.9(5.0)\binom{+12.1}{-1.8}$ MeV~\cite{HALQCD:2014okw}. In 2019, the HAL QCD Collaboration updated their investigations of $N\Omega$ system with the improved quark masses, i.e., $m_{\pi}\simeq 146$ MeV and $m_{K}\simeq 525$ MeV. Their results showed that the binding energy of $N\Omega$ bound state was $2.46(0.34) \binom{+0.04}{-0.11}$ MeV~\cite{HALQCD:2018qyu}. 

From the experimental perspective, the STAR Collaboration first measured the proton-$\Omega$ interaction correlation function in heavy-ion collisions at $\sqrt{s_{NN}}=200$ GeV in 2018~\cite{STAR:2018uho}. Their analysis slightly favored the existence of a proton-$\Omega$ bound state with a binding energy around 27 MeV. Later, the ALICE Collaboration investigated the proton-$\Omega$ strong interaction in $pp$ collisions at $\sqrt{s}=13$ TeV~\cite{ALICE:2020mfd}, but the depletion in the correlation function owing to the presence of a $p\Omega$ bound state, was not observed in the measured correlation.
 
In addition to the mass spectrum and decay properties, researches on the productions of $d_{N\Omega}$ have also been performed. Considering the high energy kaon beams at J-PARC hadron facility~\cite{Aoki:2021cqa}, COMPASS~\cite{Nerling:2012er}, OKA@U-70~\cite{Obraztsov:2016lhp}, and SPS@CERN~\cite{Velghe:2016jjw}, the $d_{N\Omega}$ production via the $K^{-} p \rightarrow d_{N \Omega}^0 \bar{\Xi}^{0}$ process was investigated in Ref.~\cite{Liu:2022uap} by using an effective Lagrangian approach, and the estimated cross sections are about several tens of nanobarn at $P_{K}=10$ GeV.  Besides the $Kp$ scattering process, one can also obtain $d_{N \Omega}$ state via the $\Omega d$ scattering process as proposed in Ref.~\cite{Aoki:2021cqa}, where the $\Omega$ beam is produced via the $K^{-}p \rightarrow \Omega^{-} \bar{K}^{(*)0} K^{+}$ process. The typical $\Omega$ beam momentum is around 3 GeV for an incident kaon momentum of 5 GeV~\cite{Aoki:2021cqa}. Moreover, as indicated in Ref.~\cite{Yuan:2021yks}, abundant hyperon beams could be accessible in the super $J/\psi$ factory, where the $\Omega$ baryon beam can be produced form the $\psi(2S) \to \Omega^- \bar{\Omega}^+ $ or $\psi(2S) \to \Omega^- K^+ \bar{\Xi}^0$ processes. Recent investigations of the (anti)hyperon-nucleon scattering carried out by the BESIII Collaboration have proved the possibilities of this scheme~\cite{BESIII:2024geh, BESIII:2023trh, BESIII:2025vqk, BESIII:2025bft}. Therefore, with the above two potential $\Omega$ beam sources, we propose to investigate the production of $d_{N \Omega}$ state in the $\Omega^{-}d \rightarrow p d_{N \Omega}$ process in the present work, where $d_{N \Omega}$ is considered as $N\Omega$ bound state with the binding energy $E_{b}=2.46$ MeV. In addition, given that $d_{N \Omega}$ predominantly decays into $\Xi \Lambda$ with a branching fraction of about $90\%$~\cite{Xiao:2020alj}, we further investigate the $\Omega^{-}d \rightarrow p \Xi \Lambda$ process and estimate the relevant cross sections.

This work is organized as follows. After Introduction, we present our estimation of the cross sections for $\Omega^{-}d \rightarrow p d_{N \Omega}^-$ and $\Omega^{-}d \rightarrow p \Xi^- \Lambda$ processes. In Section \ref{sec:MA}, the numerical results and related discussions of the cross sections are presented. The last section is devoted to a short summary.

\section{$d_{N \Omega}$ PRODUCTION IN THE $\Omega d$ scattering process}
\label{sec:MS}
	
\subsection{Cross Sections for $\Omega^{-} d \rightarrow p d_{N \Omega}^-$}

\begin{figure}[htb]
		\centering
		\includegraphics[width=75mm]{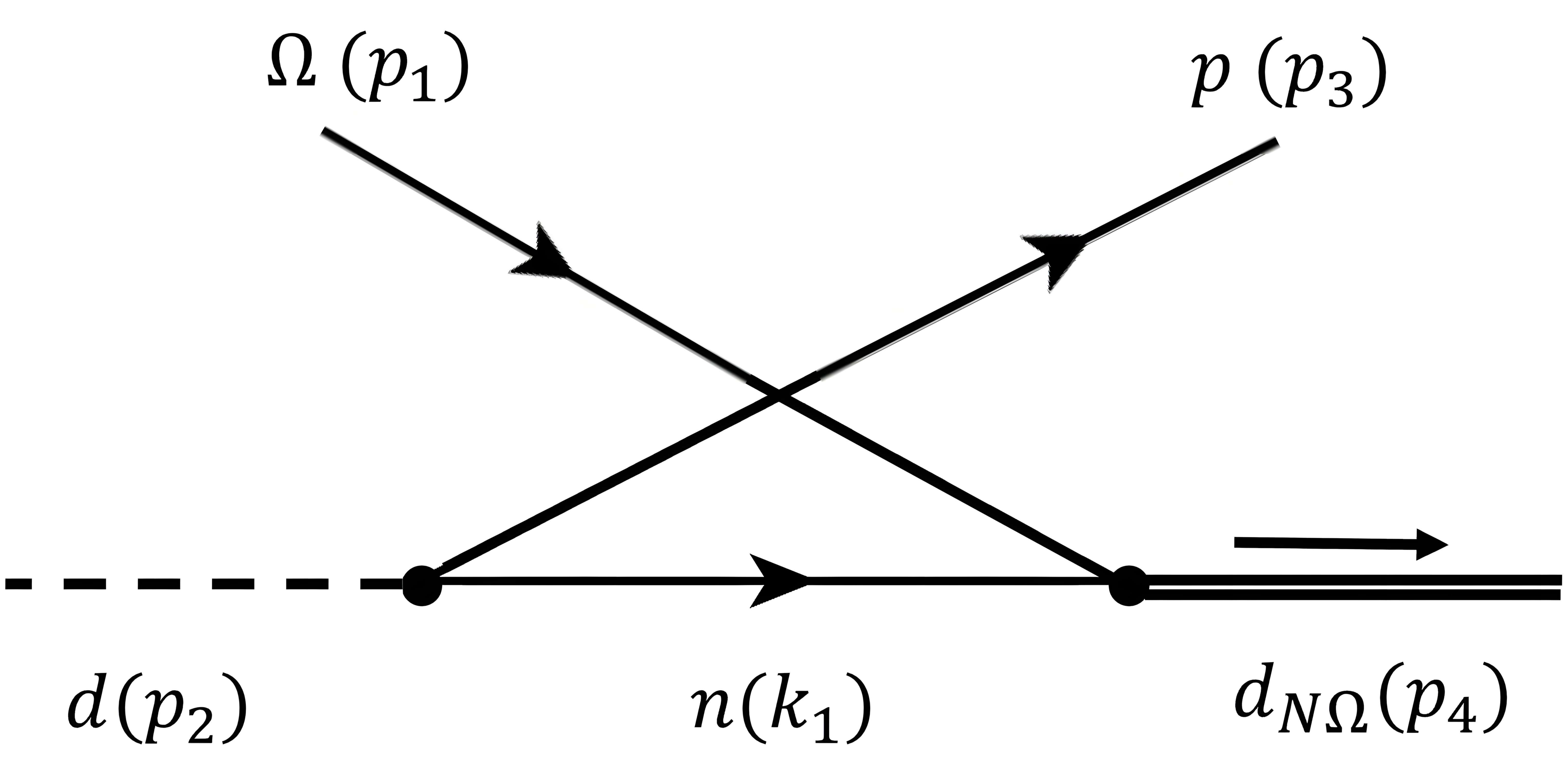}
	\caption{Diagram contributing to the $\Omega^{-} d \rightarrow p d_{N \Omega}^-$ process.}
\label{Fig.1}
\end{figure}

In this work, we consider the $d_{N \Omega}$ dibaryon to be composed of $N$ and $\Omega$ baryons with the binding energy $E_{b}=2.46$ MeV and the quantum numbers $I(J^{P})=1/2(2^{+})$. In Fig.~\ref{Fig.1}, we present the Feynman diagram for the $\Omega^{-} d \rightarrow p d_{N \Omega}$ process by the $u$-channel proton exchange. In the present estimations, we employ the effective Lagrangian approach to depict the relevant hadron interaction vertices. The effective Lagrangian for $d_{N \Omega} N \Omega$ can be written as~\cite{Xiao:2020alj},
\begin{eqnarray}
	\mathcal{L}_{d_{N \Omega} N \Omega}=g_{d_{N \Omega} N \Omega} d^{\mu \nu \dagger}_{N \Omega} \bar{\Omega}_{\mu} \gamma_{\nu} N^{c}+H.c., \label{Eq.1}
\end{eqnarray}
where $C=i \gamma^{2} \gamma^{0}$ is the charge-conjugation matrix, with the properties: $\psi^{c}=C \bar{\psi}^{T}$, $\bar{\psi}^{c}=\psi^{T} C$, $C \gamma^{\mu T}C=\gamma^{\mu}$, and $\bar{\psi}^{c}_{1} \gamma^{\mu} \psi_{2}=\bar{\psi}^{c}_{2} \gamma^{\mu} \psi_{1}$. $T$ is the transpose transformation operator. Similarly, the effective Lagrangian for $pdn$ can be written as~\cite{Weinberg:1962hj},
\begin{eqnarray}
	\mathcal{L}_{pdn}=g_{pdn} d^{\mu} \bar{p} \gamma_{5} \gamma_{\mu} n^{c} +H.c.. \label{Eq.2}
\end{eqnarray}

With the above effective Lagrangians, one can obtain the amplitude of the $\Omega^{-} d \rightarrow p d_{N \Omega}^-$ process, which is,
\begin{eqnarray}
	\mathcal{M}&=&\bar{u}^{c}(p_{3}) \Big[g_{pdn} \gamma_{5} \gamma_{\alpha}\Big] \epsilon^{\alpha}(p_{2}) S^{1/2}(k_{1},m_{p}, \Gamma_{p}) \nonumber\\ &\times& d^{\mu \nu}_{N\Omega}(p_{4}) \Big[g_{d_{N \Omega} N \Omega} \gamma_{\nu}\Big] u_{\mu}(p_{1}) \Big[F\left(k_{1},m_{p}, \Lambda_r\right)\Big]^2, \label{Eq.3}
\end{eqnarray}
where $\epsilon^{\alpha}(p_{2})$ is the polarization vector of the initial deuteron. The polarization tensor $\epsilon^{\mu \nu}(\vec{p},\lambda)$ refers to the outgoing $d_{N\Omega}$ dibaryon, and it can be constructed with the combination of the Dirac field for spin-$1/2$ and the Rarita Schwinger field for spin-$3/2$,
\begin{eqnarray}
  \epsilon^{\mu \nu}(\vec{p},\lambda)=\sum\limits_{\alpha, \,\beta} \left \langle \frac{3}{2} \alpha \frac{1}{2} \beta \big| 2 \lambda \right \rangle \psi^{\mu}_{\alpha}(\vec{p}) \gamma^{\nu} \psi^{c}_{\beta}(\vec{p}),
\end{eqnarray}
with the parameters $\lambda=(\pm2, \pm1, 0)$, $\alpha=(\pm3/2, \pm1/2)$, and $\beta=\pm1/2$, respectively. Moreover, the polarization tensor $\epsilon^{\mu \nu}$ satisfies the rules as, 
\begin{eqnarray}
p_{\mu}\epsilon^{\mu \nu}(\vec{p},\lambda)&=&p_{\nu}\epsilon^{\mu \nu}(\vec{p},\lambda)=0, \ \epsilon^{\mu \nu}(\vec{p},\lambda)=\epsilon^{\nu \mu}(\vec{p},\lambda), \nonumber\\ \epsilon^{\mu}_{\mu}(\vec{p},\lambda)&=&0, \ \epsilon^{\mu \nu *}(\vec{p},\lambda) \epsilon_{\mu \nu}(\vec{p},\lambda^{\prime})=\delta_{\lambda \lambda^{\prime}}.
\end{eqnarray}

$\mathcal{S}^{1/2}(k_{i},m_{i},\Gamma_{i})$ is the propagator of the exchanged spin-$1/2$ baryon with four-momentum $k_{i}$, mass $m_{i}$, and decay width $\Gamma_{i}$, respectively, and the detailed expression is,
\begin{eqnarray}
	\mathcal{S}^{1/2}(k_{i},m_{i},\Gamma_{i}) =\frac{{\slash\!\!\!k}_{i}+m_{i}} {k^2_{i}-m^2_{i} +i m_{i} \Gamma_{i}}.
\end{eqnarray}

The form factor $F(k_{i},m_{i},\Lambda_{r})$ is introduced to depict the inner structure of the involved hadrons in each vertex, and its specific expression is,
\begin{eqnarray}
F(k_{i},m_{i},\Lambda_{r})
=\frac{\Lambda^{4}_{r}} {\Lambda^{4} _{r} +(k^{2}_{i}-m^{2}_{i})^2},\label{Eq:Fpm2}
\end{eqnarray}
where $k_{i}$ and $m_{i}$ are the four-momentum and mass of the exchanged hadron, respectively. $\Lambda_r$ is a model parameter, and its specific value will be discussed in Section \ref{sec:MA}.

\subsection{Cross Sections for $\Omega^{-} d \rightarrow p \Xi \Lambda$}

\begin{figure}[htb]
		\centering
		\includegraphics[width=75mm]{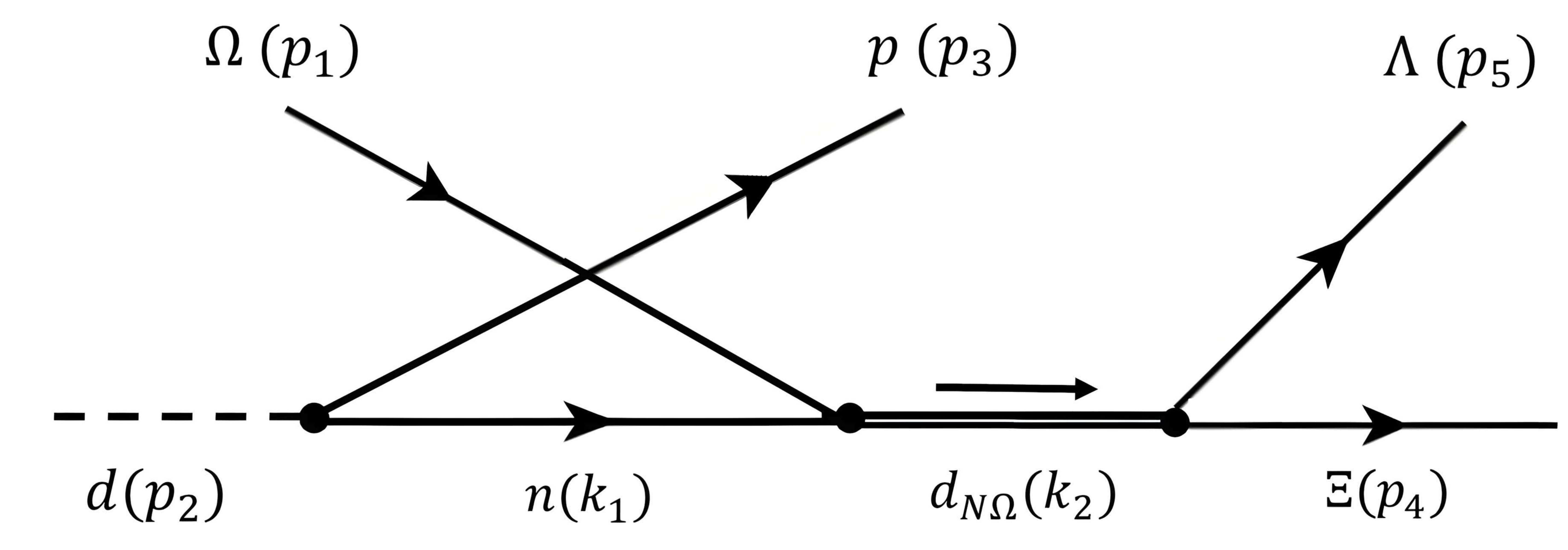}
	\caption{Diagram contributing to the $\Omega^{-} d \rightarrow p \Xi \Lambda$ process.}
\label{Fig.2}
\end{figure}

Theoretically, the authors in Ref.~\cite{Xiao:2020alj} investigated the decay properties of $d_{N \Omega}$ state, their results indicated that $d_{N \Omega}$ can decay into $\Xi \Lambda$ and $\Xi \Sigma$, respectively, while the $d_{N\Omega} \rightarrow \Xi \Lambda$ mode is dominant. Motivated by this predicted dominant decay channel, we further investigate $d_{N \Omega}$ via the $\Omega^{-} d \rightarrow p \Xi \Lambda$ process, with the corresponding Feynman diagram depicted in Fig.~\ref{Fig.2}. The effective Lagrangians for $d_{N \Omega} N \Omega$ and $pdn$ have been expressed in Eqs.~\eqref{Eq.1}-~\eqref{Eq.2}, while the effective Lagrangian for $d_{N \Omega} \Xi \Lambda$ can be written as~\cite{Liu:2022uap}, 
\begin{eqnarray}
	\mathcal{L}_{d_{N \Omega} \Xi \Lambda}&=&i \frac{g_{ d_{N \Omega} \Xi \Lambda}}{2M_{\Xi}} \bar{\Xi}^{c} \Big(\gamma_{\mu} \partial_{\nu}+\gamma_{\nu} \partial_{\mu} \Big) \Lambda d^{\mu \nu}_{N \Omega} \nonumber\\ &+& \frac{F_{d_{N \Omega} \Xi \Lambda}}{(2M_{\Xi})^2} \partial_{\mu} \bar{\Xi}^{c} \partial_{\nu} \Lambda d^{\mu \nu}_{N \Omega}+H.c.,
\end{eqnarray}
here we take $F_{d_{N \Omega} \Xi \Lambda}=0$, following the tensor dominance hypothesis~\cite{Renner:1970sbf}. With the effective Lagrangians, one can obtain the amplitude of the $\Omega^{-} d \rightarrow p \Xi \Lambda$ process, which is,
\begin{eqnarray}
\mathcal{M}^{\prime}&=&\bar{u}^{c}(p_{4}) \Big[i \frac{g_{ d_{N \Omega} \Xi\Lambda}}{2M_{\Xi}} \Big(\gamma_{\lambda}(i p^{\rho}_{5})+\gamma_{\rho} (i p^{\lambda}_{5}))\Big) \Big] u(p_{5}) \nonumber\\ &\times& S^{\mu \nu \lambda \rho}_{d_{N \Omega}}(k_{2},m_{d_{N \Omega}},\Gamma_{d_{N \Omega}}) \Big(g_{d_{N \Omega} N \Omega}  \gamma_{\nu} \Big) u_{\mu}(p_{1}) \nonumber\\ &\times& S^{1/2} \Big(k_{1},m_{p},\Gamma_{p}\Big) \Big[\bar{u}^{c}(p_{3}) \Big(g_{pdn} \gamma_{5} \gamma_{\alpha}\Big) \epsilon^{\alpha}(p_{2})\Big] \nonumber\\ &\times& \left[F\left(k_{1},m_{p},\Lambda_{r}\right)\right]^2 F\left(k_{2},m_{d_{N \Omega}}, \Lambda_{r}\right),  \label{Eq.10}
\end{eqnarray}
where $S^{\mu\nu\mu^{\prime}\nu^{\prime}}_{d_{N\Omega}}(k_2,m_{d_{N\Omega}},{\Gamma}_{d_{N\Omega}})$ is the propagator of the $d_{N\Omega}$ dibaryon, and the concrete form is,
\begin{eqnarray}
	S^{\mu\nu\mu^{\prime}\nu^{\prime}}_{d_{N\Omega}}(k_2,m_{d_{N\Omega}},{\Gamma}_{d_{N\Omega}})=\frac{i}{k^{2}_{2}-m^{2}_{d_{N\Omega}}+im_{d_{N\Omega}} {\Gamma}_{d_{N\Omega}}} \nonumber\\ \times \Big[\frac{1}{2} \Big(\tilde{g}^{\mu \mu^{\prime}} \tilde{g}^{\nu \nu^{\prime}}+\tilde{g}^{\mu \nu^{\prime}} \tilde{g}^{\nu \mu^{\prime}}\Big)- \frac{1}{3} \tilde{g}^{\mu \nu}\tilde{g}^{\mu^{\prime} \nu^{\prime}} \Big], 
\end{eqnarray}
with $\tilde{g}^{\mu \nu}=g^{\mu \nu}-k^{\mu} k^{\nu}/k^{2}$. In addition, the expression of form factor $F\left(k_{2},m_{d_{N \Omega}}, \Lambda_{r}\right)$ is same as the one in Eq.~\eqref{Eq:Fpm2}.

\section{NUMERICAL RESULTS AND DISCUSSIONS}
\label{sec:MA}

\subsection{Form factor and Coupling Constants}

The parameter $\Lambda_{r}$ in the form factor is model-dependent and is typically constrained by experimental data. However, there are no theoretical or experimental studies on the $\Omega^{-} d \rightarrow p d_{N \Omega}^-$ and $\Omega^{-} d \rightarrow p \Xi \Lambda$ processes at present. In Ref.~\cite{Jackson:2015dva}, the authors proposed that the parameter $\Lambda_{r}=0.9$ GeV for the meson-baryon-baryon vertex, so we take $\Lambda_{r}=1.0$ GeV as the central value in this work, and vary it from 0.8 to 1.2 GeV to check the parameter dependence of the cross sections. 

In addition, the values of relevant coupling constants should be clarified. In the present estimations, deuteron and $d_{N\Omega}$ are considered as the $pn$ and $N \Omega$ bound states, respectively. Thus, the coupling constants $g_{pdn}$ and $g_{d_{N \Omega} N \Omega}$ could be determined by the compositeness condition~\cite{Weinberg:1962hj, Baru:2003qq, Lin:2017mtz}. Specifically, the coupling constant $g_{pdn}$ is given by,
\begin{eqnarray}
\frac{|g_{pdn}|^2}{4 \pi}=(1-Z_{d}) \frac{(m_{p}+m_{n})^{5/2}}{(m_{p}m_{n})^{3/2}} \left(\frac{E_{b}}{2}\right)^{1/2}, \label{Eq.c}
\end{eqnarray}
where $m_{p}$ and $m_{n}$ are the masses of proton and neutron, respectively. $Z_{d}$ is a parameter for judging the properties of deuteron, while $Z_{d}=0$ denotes that there is no probability to find the deuteron as a bare (structureless) state. $E_{b}$ is the binding energy of deuteron with the value to be 2.22 MeV. Therefore, one can obtain $g_{pdn}=1.56$. Similarly, for the coupling constant $g_{d_{N \Omega} N \Omega}$, since we consider that $d_{N \Omega}$ is composed of $N$ and $\Omega$ baryons with the binding energy to be 2.46 MeV, one can also obtain $g_{d_{N \Omega} N \Omega}=1.57$ in the same way. 

The coupling constant $g_{d_{N \Omega} \Xi \Lambda}$ can be determined by combining the effective Lagrangian in Eq.~\eqref{Eq.1} and the decay width of $d_{N \Omega}\to \Xi\Lambda$. Firstly, one can obtain the amplitude $\mathcal{M}_{d_{N \Omega}\to \Xi\Lambda}$ by using an effective Lagrangian approach. Then, the decay width of the $d_{N \Omega}^-\to \Xi^-\Lambda$ process can be written as,
\begin{eqnarray}
\Gamma_{d_{N \Omega}\to \Xi\Lambda} = \frac{1}{(2J+1)8\pi} \frac{|\vec{k}_f|}{M^{2}} \overline{|\mathcal{M}_{d_{N \Omega}\to \Xi\Lambda}|^2} \label{Eq:2BDecay}
\end{eqnarray}
where $M$ and $J$ are the mass and angular momentum of the initial $d_{N \Omega}$, respectively. $\vec{k}_f$ refers to the three-momentum of the final states in the initial rest frame. Moreover, the decay properties of $d_{N \Omega}$ have been investigated in Ref.~\cite{Xiao:2020alj}, and the branching fraction of $\Xi \Lambda$ channel for $d_{N \Omega}$ is estimated to be,
\begin{eqnarray}
&&\mathcal{B}(d_{N \Omega} \rightarrow \Xi \Lambda) \simeq 90 \%.
\end{eqnarray}

Here we take the decay width of $d_{N \Omega}$ to be about $380$ keV~\cite{Xiao:2020alj}. With the above branching fraction and the decay width, one can obtain the coupling constant $g_{d_{N \Omega} \Xi \Lambda}=0.48$.

\begin{figure}[htb]
		\centering
		\includegraphics[width=85mm]{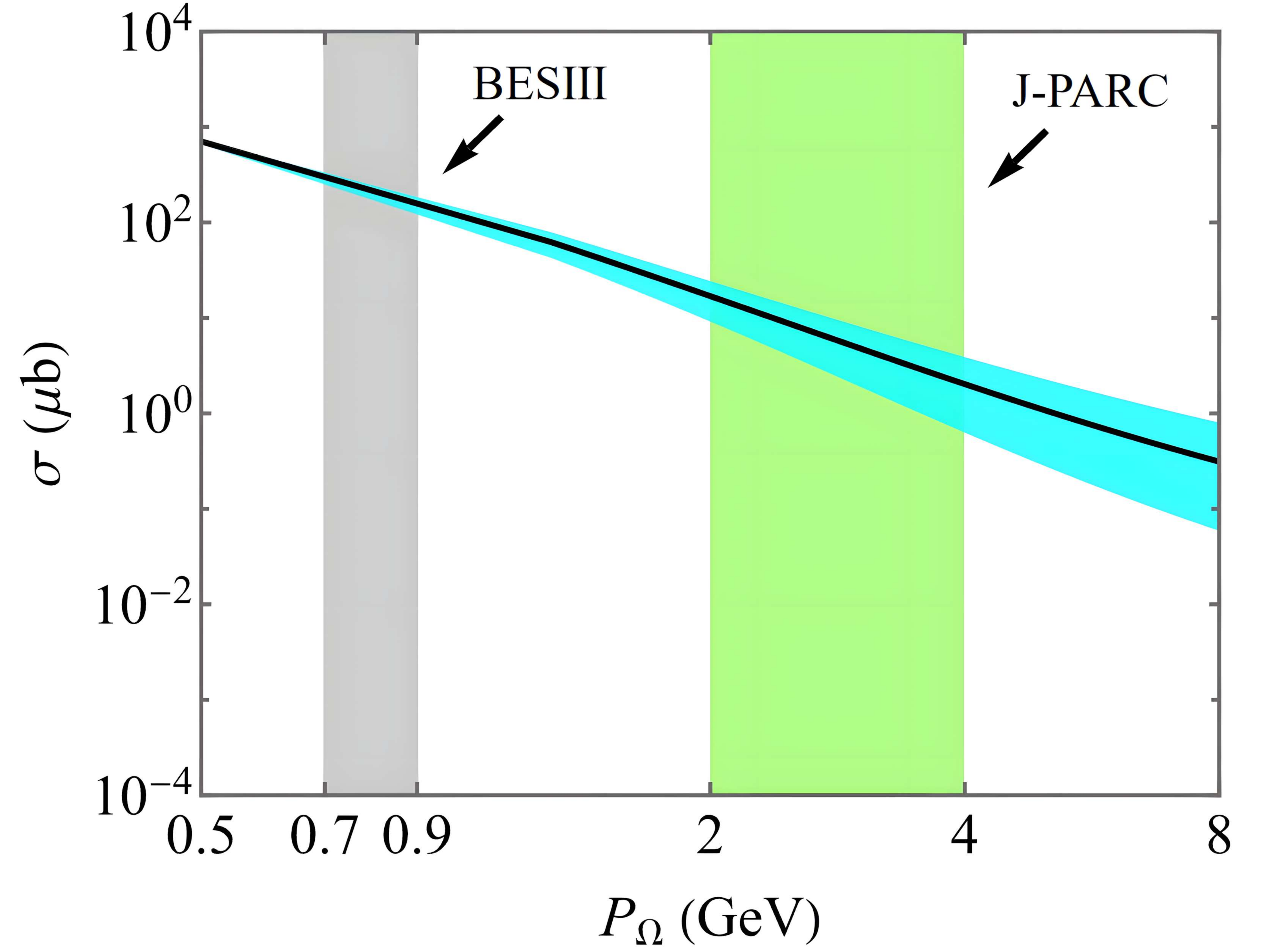}
	\caption{The cross sections for $\Omega^{-} d \rightarrow p d_{N \Omega}^-$ depending on the momentum of the incident $\Omega$ beam. The black solid curve is obtained with $\Lambda_{r}=1.0$ GeV, while the cyan band are the uncertainties resulted from the variation of $\Lambda_{r}$ from 0.8 to 1.2 GeV. The gray and green vertical bands correspond to the ranges of $P_{\Omega}=[0.7,0.9]$ GeV and $P_{\Omega}=[2,4]$ GeV, respectively.}
\label{Fig.3}
\end{figure}

\begin{figure*}[htb]
		\centering
		\includegraphics[width=175mm]{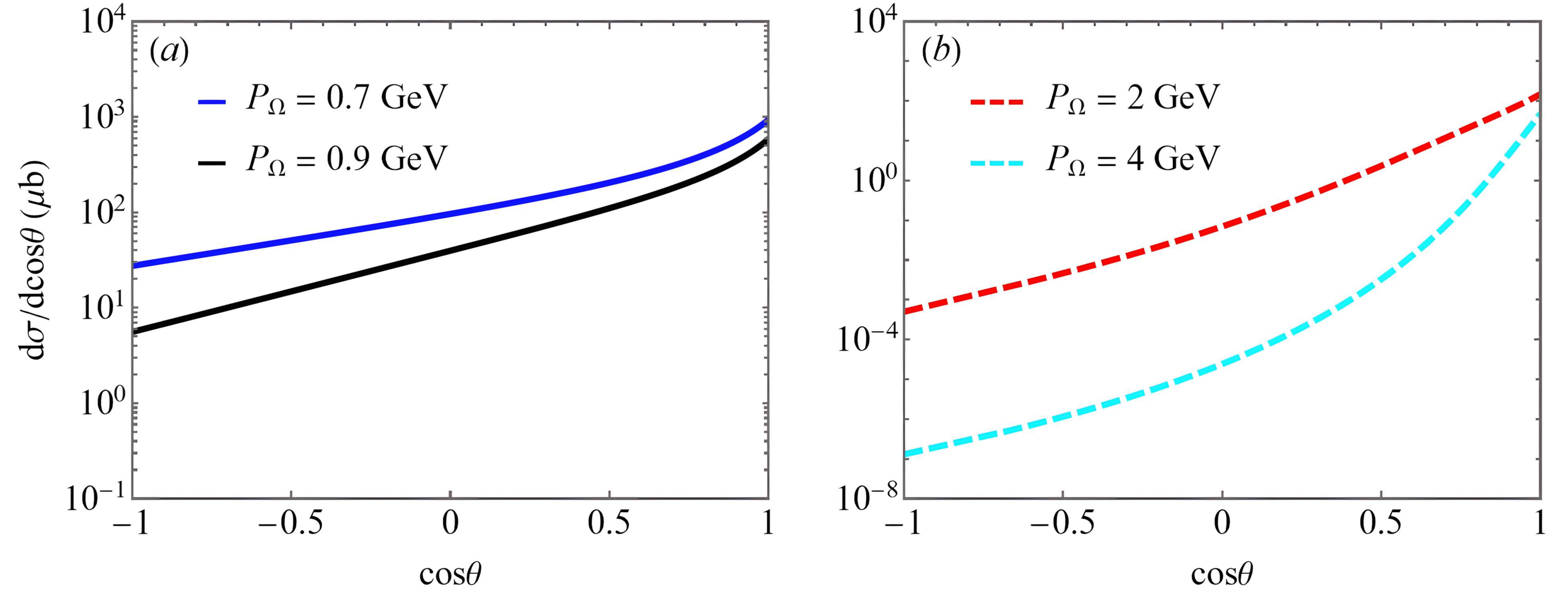}
	\caption{The differential cross section for $\Omega^{-} d \rightarrow p d_{N \Omega}^-$ depending on cos$\theta$, and the parameter $\Lambda_{r}$ is taken as 1.0 GeV.}
\label{Fig.4}
\end{figure*}

\subsection{Cross Sections for $\Omega^{-}d \rightarrow p d_{N \Omega}^-$}

After determining the parameter $\Lambda_{r}$ and relevant coupling constants, the cross sections for $\Omega^{-} d \rightarrow p d_{N \Omega}^-$ can be estimated. With the amplitude in Eq.~\eqref{Eq.3}, the differential cross sections depending on cos$\theta$ of the $\Omega^{-} d \rightarrow p d_{N \Omega}^-$ process can be written as,
\begin{eqnarray}
	\frac{d{\sigma}} {d \cos\theta}
	=\frac{1} {32 \pi s} \frac{|\vec{p}_f|} {|\vec{p}_i|}  \left(\frac{1} {12} \left|\overline{\mathcal{M}} \right|^2\right),
\end{eqnarray}
where $s$ and $\theta$ refer to the square of the center-of-mass energy and the scattering angle, respectively, while $\theta$ is the angle between the outgoing $d_{N \Omega}$ and the initial $\Omega$ beam direction in the center-of-mass system. The $\vec{p}_f$ and $\vec{p}_i$ refer to the three momenta of the outgoing $d_{N \Omega}$ and the initial $\Omega$ beam in the center-of-mass system, respectively. 

With the above preparations, the cross sections for the $\Omega^{-} d \rightarrow p d_{N \Omega}^-$ process depending on the momentum of the incident $\Omega$ beam are presented in Fig.~\ref{Fig.3}. The black solid curve corresponds to the cross sections obtained with $\Lambda_{r}=1.0$ GeV, and the cyan band indicates the uncertainties resulted from the variation of $\Lambda_{r}$ from 0.8 to 1.2 GeV. Moreover, the gray and green vertical bands correspond to the ranges of $P_{\Omega}=[0.7,0.9]$ GeV and $P_{\Omega}=[2,4]$ GeV, respectively, which are around the momentum of $\Omega$ beam come from the $\psi(2S) \rightarrow \Omega^{-} \bar{\Omega}^{+}$ decay process at BES\uppercase\expandafter{\romannumeral3}~\cite{Yuan:2021yks} and the $K^{-}p \rightarrow \Omega^{-} K^{+} \bar{K}^{(*)0}$ process at J-PARC~\cite{Aoki:2021cqa}, respectively. From the figure, one can find that the cross sections decrease gradually with the increasing of incident $\Omega$ beam. At $P_{\Omega} =$ 0.7, 0.9, 2.0, and 4.0 GeV, the cross sections for $\Omega^{-} d \rightarrow p d_{N \Omega}^-$ are $\Big(329.7^{+26.9}_{-49.6}\Big)$ $\mu$b, $\Big(174.0^{+26.5}_{-38.2}\Big)$ $\mu$b, $\Big(16.9^{+7.4}_{-7.7}\Big)$ $\mu$b, and $\Big(2.0^{+1.8}_{-1.4}\Big)$ $\mu$b, respectively, where the central values are estimated with $\Lambda_{r}=1.0$ GeV, and the errors come from the variation of $\Lambda_{r}$ from 0.8 to 1.2 GeV. In addition, one can find that the cross sections for $\Omega^{-} d \rightarrow p d_{N \Omega}^-$ range from several hundreds nanobarn to several hundred microbarn as $\Lambda_{r}$ varies in the range of $P_{\Omega}=[0.7,4.0]$ GeV.

In addition to the cross sections, the differential cross sections for $\Omega^{-} d \rightarrow p d_{N \Omega}^-$ depending on cos$\theta$ are presented in Fig.~\ref{Fig.4}, where $\theta$ is the angle between the outgoing $d_{N \Omega}$ and the incident $\Omega$ beam direction. It is worth noting that we take the parameter $\Lambda_{r}=1.0$ GeV in the present estimations. The blue solid, black solid, red dashed, and cyan dashed curves stand for the differential cross sections at $P_{\Omega}=0.7$, 0.9, 2.0, and 4.0 GeV, respectively. Our estimations show that the differential cross sections all reach the maximum at the forward angle limit. Furthermore, more $d_{N\Omega}$ states are concentrated in the forward angle area as $P_{\Omega}$ increases.

\subsection{Cross sections for $\Omega^{-} d \rightarrow p \Xi \Lambda$}

\begin{figure}[htb]
		\centering
		\includegraphics[width=85mm]{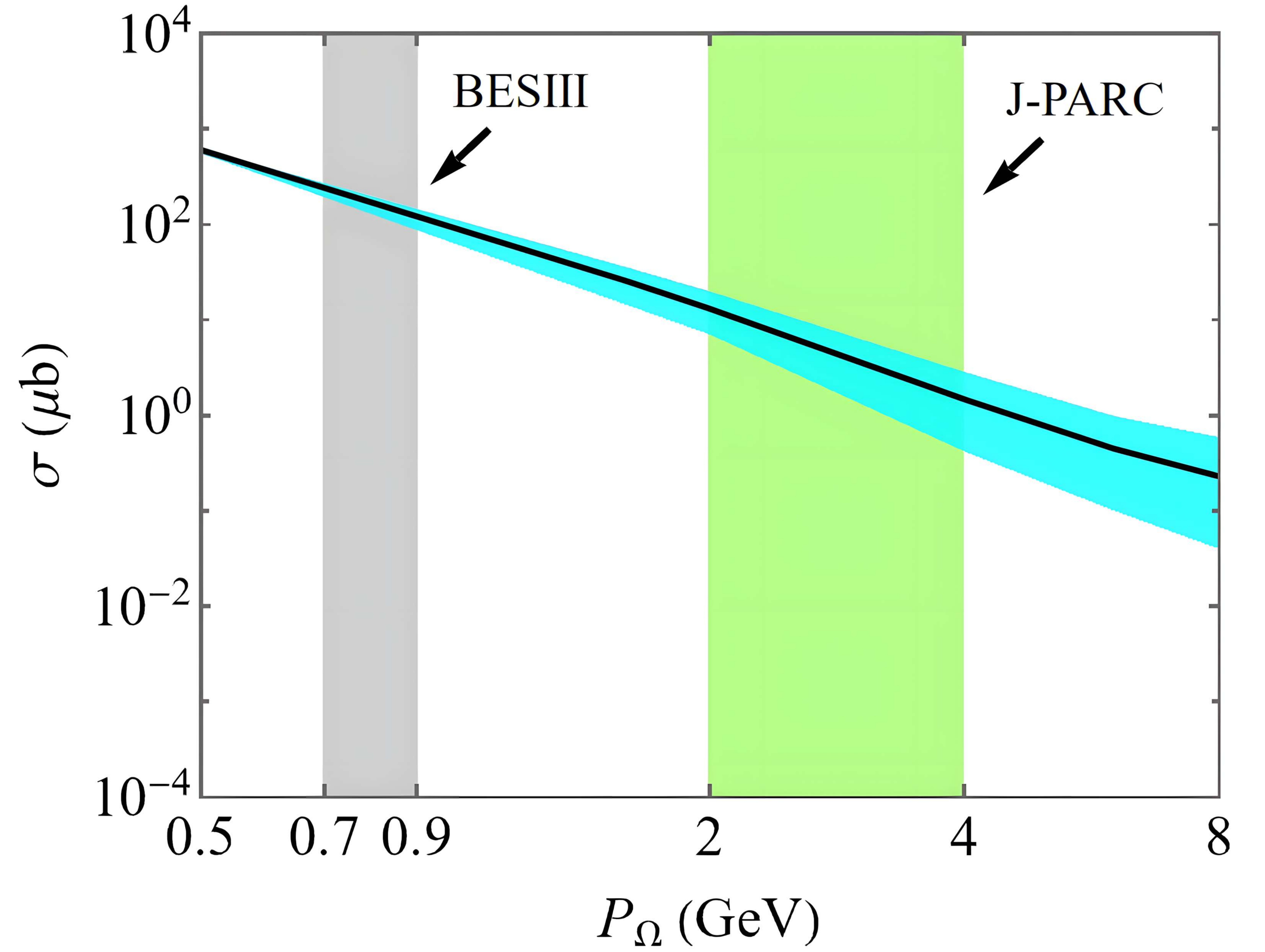}
	\caption{The cross sections for the $\Omega^{-} d \rightarrow p \Xi \Lambda$ process depending on the momentum of the incident $\Omega$ beam. The black solid curve is obtained with $\Lambda_{r}=1.0$ GeV, while the cyan band are the uncertainties resulted from the varying of $\Lambda_{r}$ from 0.8 to 1.2 GeV. The gray and green rectangles correspond to the ranges of $P_{\Omega}=[0.7,0.9]$ GeV and $P_{\Omega}=[2,4]$ GeV, respectively.}
\label{Fig.5}
\end{figure}

Since the $d_{N \Omega}$ state decays predominantly into $\Xi \Lambda$, we further estimate the cross sections for the $\Omega^{-} d \rightarrow p \Xi \Lambda$ process. With the amplitude given in Eq.~\eqref{Eq.10}, the differential cross section of the $\Omega^{-} d \rightarrow p \Xi \Lambda$ process can be written as,
\begin{eqnarray}
	d{\sigma}=\frac{1} {8(2 \pi)^4} \frac{1} {\Phi} \left(\frac{1} {12} \left|\overline{\mathcal{M^{\prime}}} \right|^2\right) d p^0 _5 d p^0 _3 d\cos\theta d \eta,
\end{eqnarray}
where the flux factor $\Phi=4|{\vec{p_1}}|\sqrt{s}$, $\vec{p_1}$ and $\sqrt{s}$ stand for the three-momentum of the initial $\Omega^{-}$ hyperon and the center-of-mass energy, respectively. $p^0_3$ and $p^0_5$ are the energy of the outgoing neutron and $\Lambda$, respectively.

The cross sections for $\Omega^{-} d \rightarrow p \Xi \Lambda$ depending on the momentum of the incident $\Omega$ beam are presented in Fig.~\ref{Fig.5}, where the black curve is estimated with $\Lambda_{r}=1.0$ GeV, and the cyan band indicates the uncertainties resulted from $\Lambda_{r}$. The gray and green vertical bands correspond to the ranges of $P_{\Omega}=[0.7,0.9]$ GeV and $P_{\Omega}=[2,4]$ GeV, respectively. From the results, one can find that the cross sections decrease gradually with the increasing of incident $\Omega$ beam. In particular, the cross sections for the $\Omega^{-} d \rightarrow p \Xi \Lambda$ process at $P_{\Omega} =$ 0.7, 0.9, 2.0, and 4.0 GeV are $\Big(299.4^{+25.7}_{-42.8}\Big)$ $\mu$b, $\Big(155.7^{+21.8}_{-34.6}\Big)$ $\mu$b, $\Big(13.0^{+6.9}_{-6.0}\Big)$ $\mu$b, and $\Big(1.5^{+1.3}_{-1.1}\Big)$ $\mu$b, respectively. Moreover, a comparison of the cross sections in Figs.~\ref{Fig.3} and ~\ref{Fig.5} shows that the overall cross sections for $\Omega^{-} d \rightarrow p \Xi \Lambda$ are about $90\%$ of those for $\Omega^{-}d \rightarrow p d_{N \Omega}^-$, which is well consistent with the branching fraction of $d_{N \Omega}^- \rightarrow \Xi^- \Lambda$.

Experimentally, the BEPCII storage ring~\cite{Yu:2016cof} has enabled the BESIII experiment to accumulate extensive data samples from $e^{+}e^{-}$ collisions, including 10 billion $J/\psi$ and 3 billion $\psi(2S)$ events~\cite{BESIII:2020nme, Lu:2020imt, Zhang:2022bdc, Zhang:2026qjt}. With $\mathcal{B}[\psi(2S)\to \Omega^- \bar{\Omega}^+]=(5.66\pm 0.30)\times 10^{-5}$, the events of $\Omega^-$ baryon  produced at BESIII is $1.8\times 10^{5}$. For a liquid deuteron target, the density of deuteron is about $4.8 \times 10^{22}/\mathrm{cm}^{-3}$. The $\Omega$ baryon with $P_{\Omega}=700$ MeV can fly more than 1 cm  in its mean life, thus the equivalent surface density of the deuteron is about $4.8 \times 10^{22}/\mathrm{cm}^{-2}$. Together with $\sigma(\Omega^- d \to p d_{N\Omega}^-)=300 \ \mathrm{\mu b}$ estimated in the present work, we can conclude only several $d_{N\Omega}^-$ could be produced with the present $\psi(2S)$ data sample. It is worth noting that the Super Tau-Charm Facility (STCF) is designed to operate with a center-of-mass energy range of 2 to 7 GeV and a peak luminosity of 5.0 $\times$ $10^{34}$ $\mathrm{cm}^{-2}\mathrm{s}^{-1}$, which translates to a yield of events more than 50 times greater than that of BEPCII~\cite{Peng:2020orp, Achasov:2023gey, Ai:2025xop}. With the addition of a dedicated high-density fixed deuteron target, our estimates suggest an event yield on the order of $10^{2}$ at such a facility. For the J-PARC facility, the $\Omega$ beam could be produced via the $K^{-}p \rightarrow \Omega^{-} \bar{K}^{(*)0} K^{+}$ scattering. However, the lack of specific cross-section data for this process currently prevents an effective prediction of the number of events.

\section{SUMMARY}
\label{sec:NR}

Theoretically, the $d_{N \Omega}$ state is generally described as a bound, six-quark configuration with strangeness $S=-3$. In recent years, researches on $d_{N \Omega}$ have attracted increasing attentions. Among them, the production properties of $d_{N \Omega}$ are important for the explorations of its inner structure. On the one hand, the STAR Collaboration has investigated the proton-$\Omega$ interaction correlation function in heavy-ion collisions at $\sqrt{s_{NN}}=200$ GeV. Moreover, the ALICE Collaboration investigated the proton-$\Omega$ strong interaction in $pp$ collisions at $\sqrt{s}=13$ TeV. On the other hand, with the high energy kaon beam provided from the J-PARC hadron facility, COMPASS, OKA@U-70, and SPS@CERN, the $K^{-} p$ scattering process was proposed to investigate the production of $d_{N \Omega}$. 

In addition to the above mentioned processes, one can also investigate the production of $d_{N \Omega}$ via the $\Omega d$ scattering process if the initial $\Omega$ beam could be obtained. Experimentally, the J-PARC hadron facility proposed to investigate the $K^{-}p \rightarrow \Omega^{-} \bar{K}^{(*)0} K^{+}$ process, their results indicated that the typical momentum of $\Omega$ beam is around 3 GeV. Theoretically, based on the 3 billion $\psi(2S)$ event data sample at BESIII, the $\psi(2S) \rightarrow \Omega^{-} \bar{\Omega}^{+}$ or $\psi(2S) \to \Omega^- K^+ \bar{\Xi}^0$ decay processes have also been studied, and the results showed that the maximum momentum of $\Omega$ beam is about 774 MeV. Therefore, considering these two potential $\Omega$ beam sources, we propose to investigate the production of $d_{N \Omega}$ in the $\Omega^{-}d \rightarrow p d_{N \Omega}^-$ process in this work. 

For the $\Omega^{-} d \rightarrow p d_{N \Omega}^-$ process, our estimations indicate that the cross sections decrease gradually with the increasing of incident $P_{\Omega}$. In particular, the cross sections at $P_{\Omega} =$ 0.7, 0.9, 2.0, and 4.0 GeV are $\Big(329.7^{+26.9}_{-49.6}\Big)$ $\mu$b, $\Big(174.0^{+26.5}_{-38.2}\Big)$ $\mu$b, $\Big(16.9^{+7.4}_{-7.7}\Big)$ $\mu$b, and $\Big(2.0^{+1.8}_{-1.4}\Big)$ $\mu$b, respectively, where the central values are estimated with $\Lambda_{r}=1.0$ GeV, and the errors come from the variation of the parameter $\Lambda_{r}$ from 0.8 to 1.2 GeV. Moreover, the differential cross sections for $\Omega^{-} d \rightarrow p d_{N \Omega}^-$ are also estimated in this work. Since the $d_{N \Omega}$ state can decay into $\Xi \Lambda$ and $\Xi \Sigma$, and the $d_{N \Omega} \rightarrow \Xi \Lambda$ decay mode is dominant. We further estimate the cross sections for the $\Omega^{-} d \rightarrow p \Xi \Lambda$ process. At $P_{\Omega} =$ 0.7, 0.9, 2.0, and 4.0 GeV, our estimations show that the cross sections are $\Big(299.4^{+25.7}_{-42.8}\Big)$ $\mu$b, $\Big(155.7^{+21.8}_{-34.6}\Big)$ $\mu$b, $\Big(13.0^{+6.9}_{-6.0}\Big)$ $\mu$b, and $\Big(1.5^{+1.3}_{-1.1}\Big)$ $\mu$b, respectively. It is expected that the present results can be tested by further experimental measurements at J-PARC and STCF in the future.

\section*{ACKNOWLEDGMENTS}
This study is partly supported by the National Natural Science Foundation of China under Grant Nos. 12175037 and 12335001, and is supported, in part, by National Key Research and Development Program under contract No. 2024YFA1610503. Quan-Yun Guo is supported by the SEU Innovation Capability Enhancement Plan for Doctoral Students (Grant No. CXJH SEU 26160)


\end{document}